\documentclass[fleqn,twoside]{article}
\usepackage{espcrc2}

\usepackage{graphicx}

\usepackage[figuresright]{rotating}

\newcommand{\AmS}{{\protect\the\textfont2
  A\kern-.1667em\lower.5ex\hbox{M}\kern-.125emS}}

\title{\LARGE{\bf Physics of the CLAS collaboration: selected results}}

\author{Patrizia Rossi (on behalf of the CLAS Collaboration)\\
 INFN Laboratori Nazionali di Frascati, 
        via E. Fermi 40 I-00044; Frascati, Italy\\
	}
\begin{document}
\pagestyle{empty}

\begin{abstract}
{\it The CLAS collaboration in Hall B at the Thomas Jeffeson National Accelerator Facility (Jefferson Lab) has a broad scientific program. Recent results on the main topics are presented. These concern the electromagnetic exitation of nucleon resonances, the measurement of inclusive spin structrure functions in the nucleon resonance region, the first signature of the generalized parton distributions through deeply virtual Compton scattering measurements, deuteron photodisintegration and the photoproduction of light vector mesons.}
\end{abstract}

\maketitle
\section{INTRODUCTION}
The superconducting electron accelerator at Jefferson Lab delivers a low-emittance, high resolution, 100\% duty-cycle electron beam to three different experimental Halls A, B and C, simultaneously. The maximum energy is at least 5.7, and closer to 5.8 GeV (with 80\% polarization available) and the maximum current is 180 $\mu$A.
\\
Hall B is devoted to experiments that require the detection of several particles in the final state. The CLAS detector \cite{clas} is unique in design to accomodate these requirements and to perform a very broad spectrum of physics measurements.

The themes of the Hall B scientific program are the precision study of the structure of the nucleon and the nature of the strong interaction. Experiments aim to clarify the interplay between hadronic and partonic degrees of freedom and the effectiveness of the traditional nucleon-nucleon theories or QCD inspired models.
\\
This scientific program can be summarized in the following main topics:
\begin{itemize} 
\item {\it Baryon Resonances}
\item {\it Spin Structure Functions in the Resonance Region}
\item {\it Nucleon Tomography}
\item {\it Dynamics of the Strong Interactions}
\end{itemize}
\section{BARYON RESONANCES}
The study of the baryon spectrum is a fundamental part of the scientific program in Hall B. The so-called $N^\ast$ program is concerned with the electromagnetic production of exclusive hadronic final states, with the purpose of extracting information on excited baryon states that will reveal underlying symmetries and internal dynamics.

A lot of progress has been made in the framework of quark models to provide a consistent picture of all resonant states in terms of three constituent, massive quarks. Various recipes have been introduced to account for relativity and to reproduce form factors \cite{capstick,cardarelli}, while other models rely on a different treatment of the basic degrees of freedom and the potential \cite{santopinto}.

 It is the goal of the CLAS experimental project to provide accurate data to test these different quark-model approaches to nucleon structure.

In the following, I will show some selected issues in the $N^\ast$ program both in single pion electroproduction, suitable for investigating the first and second resonance regions, and multiple pion electroproduction.
\subsection{Quadrupole Deformations of the Nucleon and the Delta Resonance}
An interesting aspect of nucleon structure at low energies is a possible quadrupole deformation of the nucleon and the $\Delta(1232)$.
Within SU(6) models, the $\gamma^\ast N \rightarrow \Delta(1232) \rightarrow N\pi$ transition is mediated by a single quark spin flip, leading to $M_{1+}$ dominance and $E_{1+}$=$S_{1+}$= 0. Non-zero values for $E_{1+}$ would indicate the quadrupole deformation. Such deformation may dynamically arise through the interaction of the photon with the pion cloud \cite{sato,kamalov} or through the one-gluon exchange mechanism \cite{koniuk}. Finally, quark helicity conservation in perturbative QCD (pQCD) requires $E_{1+}$=$M_{1+}$ as $Q^2 \rightarrow \infty$ \cite{warren}.
Determination of the ratios $R_{EM}= E_{1+}/M_{1+}$ and $R_{SM}= S_{1+}/M_{1+}$ has been the aim of a considerable number of experiments in the past, but they suffered from large systematic and statistical errors, as well as a limited kinematic coverage.

CLAS, with its nearly full coverage in azimuthal and polar angles, allows the extraction of the multipoles with greater accuracy than in the past.
Results of the multipole analysis of the CLAS data \cite{joo1} for the reaction $ ep \rightarrow e'p\pi^0 $ are shown in Fig.1, where data from previous experiments published after 1990 are included as well \cite{beck,blanpied,frolov}. As one can see $R_{EM}$ remains negative and small throughout the $Q^2$ range. There are no indications that leading pQCD contributions are important. The ratio $R_{SM}$ also remains negative, but its magnitude is strongly rising with $Q^2$. The comparison with microscopic models, relativized quark models \cite{warns,azn}, chiral quark soliton model \cite{silva}, dynamical models \cite{sato,kamalov,kamalov2} shows that the simultaneous description of both $R_{EM}$ an $R_{SM}$ is achieved by dynamical models that include the nucleon pion cloud.

\begin{figure}[ht]
\vspace{20pt}
\includegraphics[width=8.cm, height=9.cm]{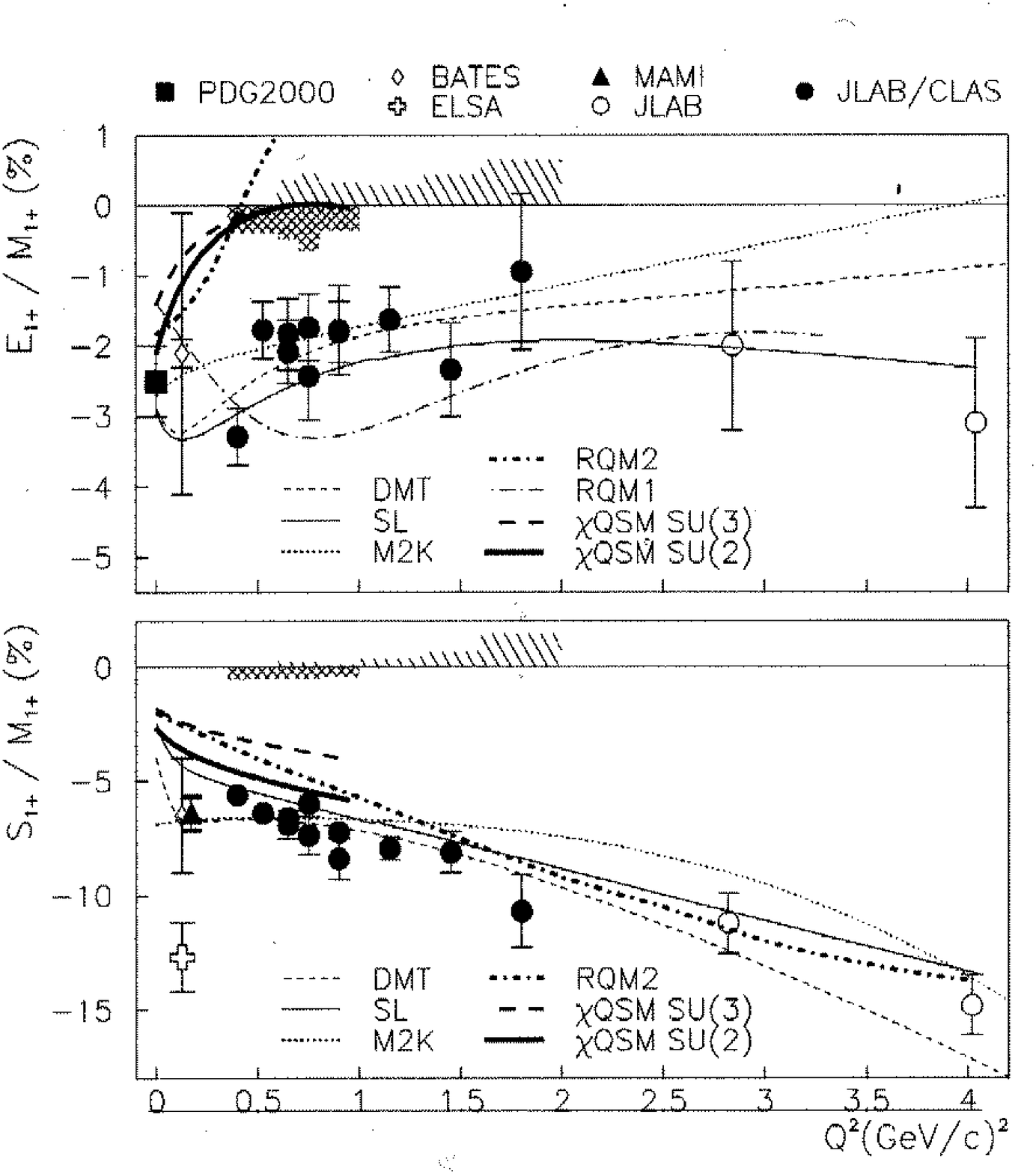}
\vspace{-1.2cm}
\caption{$R_{EM}$ and $R_{SM}$ from CLAS (full circle)\cite{joo1} and from the experiments after 1990: BATES\cite{mertz}, ELSA\cite{kalle}, JLAB (Hall C)\cite{frolov} and MAMI\cite{mami}. The curves show recent model calculations (see text): $\chi$QSM\cite{silva}, DMT\cite{kamalov}, SL\cite{sato}, M2K\cite{kamalov2}, RQM1\cite{warns} and RQM2\cite{azn}. }
\label{fig:fig1}
\end{figure}

Model dependencies in the analysis are largely due to the poor knowledge of the non-resonant terms, which become increasingly important at higher $Q^2$. Unfortunately, cross section measurements alone are unable to separate the $\Delta(1232)$ from the non-resonant background. Polarization observables, on the other hand, can access directly the interference between these processes.
The differential cross section for $\gamma^\ast p \rightarrow p\pi^0$ with a longitudinally polarized electron beam and an unpolarized target, depends on the transverse $\epsilon$ and longitudinal $\epsilon_L$ polarization of the virtual photon and five structure functions: $\sigma_T$, $\sigma_L$, $\sigma_{TT}$, $\sigma_{LT}$ and $\sigma_{LT'}$.
\\
The interference term $\sigma_{LT'}$ has been measured for the first time with CLAS \cite{joo2} in the nucleon resonance region using a polarized electron beam and out-of-plane for the pion. The results are shown in Fig.2 compared with dynamical models which clearly show the sensitivity to non-resonant contributions. In fact, all models predict nearly the same unpolarized cross sections at the $\Delta$ mass (upper panel for W=1.22 GeV), differing in their handling of non-resonant contributions.

\begin{figure}[ht]
\vspace{3pt}
\includegraphics[width=7.5cm, height=8.5cm]{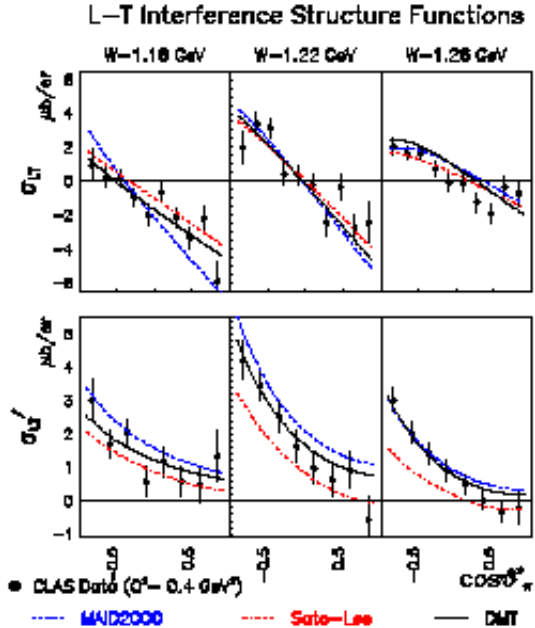}
\vspace{-1.2cm}
\caption{Response functions $\sigma_{LT}$ and $\sigma_{LT'}$ for $\pi^0$ production from protons measured with CLAS\cite{joo2} compared to predictions of three dynamical models \cite{dre,sato,kamalov}. The $\sigma_{LT'}$ data show strong sensitivity to the non-resonant contributions in the various models.}
\label{fig:fig2}
\end{figure}

\begin{figure}[ht]
\includegraphics[width=8.cm, height=9.0cm]{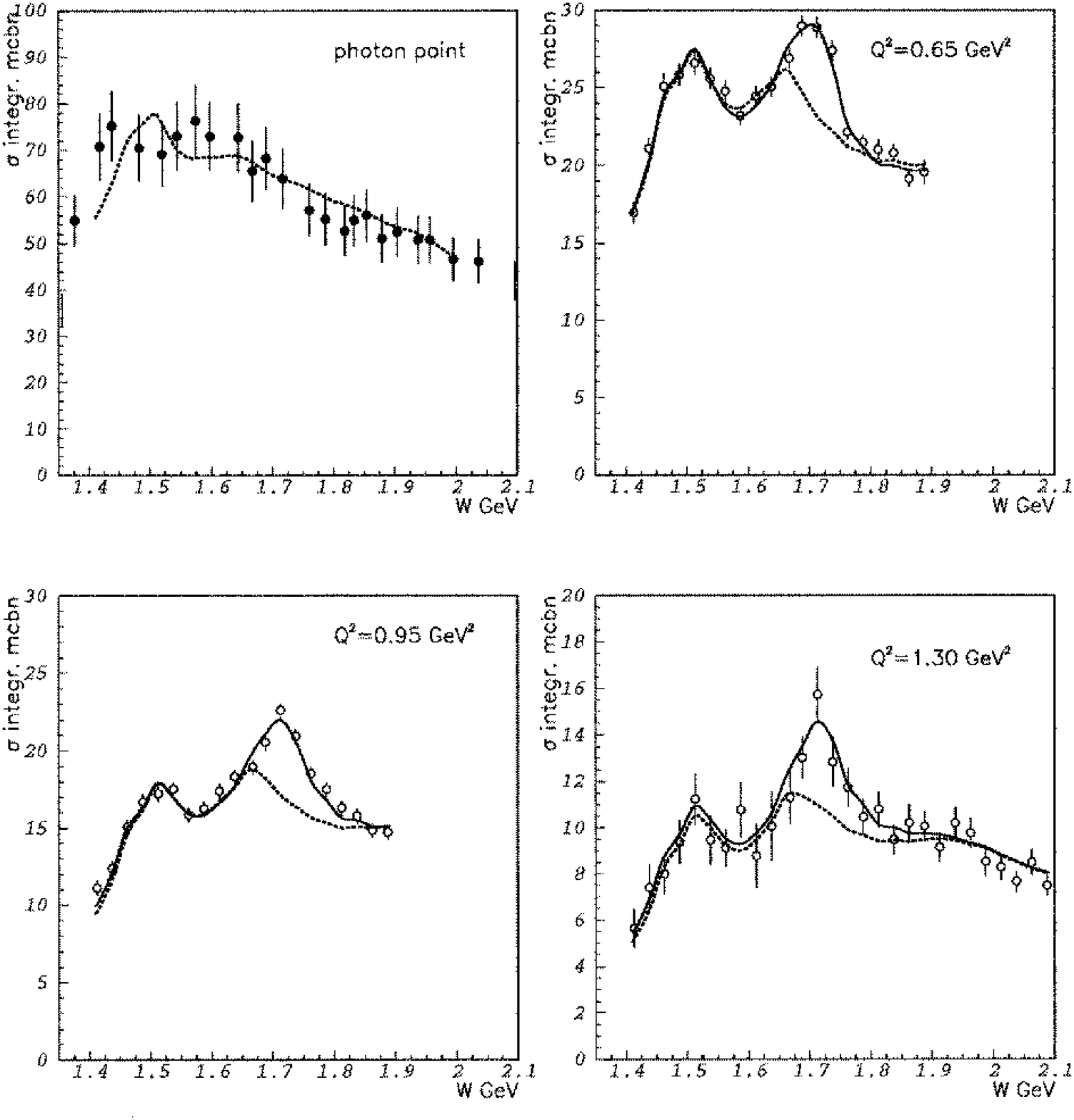}
\vspace{-1.2cm}
\caption{Total photoabsorption cross section for $\gamma^\ast p \rightarrow p\pi^+\pi^-$. Top left panel: photoproduction data from DESY. The other panel show the CLAS data at $Q^2=0.65,0.95,1.3\ \rm{GeV^2}$. For each $Q^2$ bin is evident the structure near 1.7 GeV. The dashed line represents our knoledge of $N^\ast$ electromagnetic and hadronic properties with the coupling varied within the empirical uncertainties. The solid line is a best fit to the data assuming the existance of a second $N^\ast_{{3/2}^+}$(1720) with different hadronic couplings.}
\label{fig:fig3}
\end{figure}

\subsection{Missing Resonances}
SU(6) symmetric quark models \cite{koniuk,giannini} predict more states than have been found in experiments. QCD mixing effects could decouple many of these states from the pion-nucleon channel \cite{koniuk}, with a consequent lack of evidence in elastic $\pi N$ scattering, while strongly coupling them to two-pion channels such as $\Delta\pi$ \cite{koniuk,koniuk2,capstick2}. However, other models such as the Quark Cluster Model \cite{liu} predict fewer states than the symmetric model, which is more in accordance with experimental observations.

The search for some of these states is therefore crucial for discriminating  between alternative descriptions of baryon structure.

The new CLAS total cross section electroproduction data \cite{ripani} for the reaction $e p\rightarrow e'p \pi^+ \pi^-$ are shown in Fig. 3 in comparison with photoproduction data from DESY \cite{desy}. The most striking feature is the strong resonance peak near the invariant mass W=1.72 GeV which is seen here for the first time but is absent in photoproduction data. Analyzing the complete hadronic angular distributions and $p\pi^+$ and $\pi^+\pi^-$ mass distributions over the full W range, has shown that the peak near 1.72 GeV is best described by a $N^\ast_{{3/2}^+}(1720)$ state. Although there exists a state with such quantum numbers in this mass range, its hadronic properties were previously found to be very different from the resonance seen in the CLAS data.

Therefore, this state could be one of the ``missing'' states. In fact, Capstick and Roberts \cite{capstick2} predict a second $N^\ast_{{3/2}^+}$ state at a mass 1.87 GeV and Capstick and Page \cite{capstick3} predict a hybrid baryon state with these quantum numbers at about the same mass. Mass predictions of these models are uncertain by at least $\pm 100$ MeV, and therefore are not inconsistent with the observed state at 1.72 GeV. Independent of possible interpretations, the hadronic properties of the state seen in the CLAS data appear incompatible with the known states listed in the Review of Particle Properties \cite{RPP} and deduced from $N\pi\pi$ final states in $\pi N$ scattering.

\section{SPIN STRUCTURE FUNCTIONS}
The nucleon spin structure functions, $g_1(x)$ and $g_2(x)$, in the deep-inelastic scattering (DIS) region have been widely investigated, both experimentally and theoretically, since the results of the EMC experiment \cite{EMC} performed at CERN found that the quark spins contribute little to the nucleon spin. 

Actually, much of the recent work has focused on measuring the fundamental Bjorken sum rule \cite{bj}:
\begin{equation}
\int^1_0{[g_1^p(x)-g_1^n(x)]dx} = \frac{1}{6}g_A
\end{equation}
where $g_1^p (g_1^n)$ is the spin structure function for the proton (neutron), $g_A$  is axial coupling constant, and the integration is taken over the full range of {\it x} Bjorken ({\it $x_B$}). Eq.1 only holds true at infinite. However, it has been evolved to the finite $Q^2$ using pQCD and this agrees with experiment to within $5\%$.

At low energies the validity of pQCD comes into question, and there is no way to predict or calculate accurately the spin structure functions  without directly measuring them. Experiments at very low $Q^2$ can use the Gerasimov-Drell-Hearn (GDH) sum rule \cite{GDH} (Eq.2) as a guide to study the evolution of the nucleon spin structure functions:
\begin{equation}
\int^{\infty}_{thr}[\sigma_{1/2}-\sigma_{3/2}]\frac{d\nu}{\nu} = - \frac{2\pi^2\alpha}{M^2}{\kappa^2}
\end{equation}
The GDH sum rule relates the difference in polarized cross section $\sigma_{1/2, 3/2}$, for the scattering of real photons, to the target's anomalous magnetic moment $\kappa$, mass M, and the electromagnetic coupling constant $\alpha$.

This sum rule has been studied for photon energies up to 2.5 GeV \cite{ahrens} and in this limited energy range deviates from the theoretical asymptotic values by less than $10\%$.

Recently Ji and Osborn \cite{osborne} extended the sum rule by linking the integral to the spin-dependent virtual Compton amplitude. Their work unified the two fundamental sum rules: the Bjorken sum rule in the DIS regime and the GDH sum rule at the real photon point.

A measurement of the $Q^2$-dependence allows test of the GDH sum rule evolution and chiral perturbation theory at low $Q^2$ and sheds light on the question at what distance scale pQCD corrections and the QCD twist expansion will break down, and where the physics of confinement will dominate. It will also allow one to evaluate where resonances give important contributions to the first moment of $g_1$ \cite{burkert}.

The spin structure functions experimental program at CLAS is focused just on the moderate $Q^2$ regime and in the domain of nucleon resonances. The first round of experiments has been completed on polarized hydrogen ($NH_3$) and deuterium ($ND_3$). In the following I will show the results obtained for the inclusive spin structure functions of deuteron.
\subsection{Inclusive Spin Structure Functions of the Deuteron}
Results from a new measurement of spin structure functions of the deuteron for moderate momentum transfer ($Q^2=0.27-1.3\ \rm{GeV^2}$) and final hadronic state mass in the nucleon resonance region ($W = 1.08-2.0\ \rm{GeV}$) have been obtained at CLAS with a 2.5 GeV polarized electron beam \cite{khun}.

The spin structure function $g_1^d$ was extracted from the photon asymmetry measurement for four different bins in $Q^2$ and then the integrals $\Gamma_1^d(Q^2)$ = $\int{g_1^d(x,Q^2)dx}$ were calculated.

Results are shown in Fig. 4 together with theoretical predictions and the only experimental data \cite{slac1} available before CLAS. The solid line at higher $Q^2$ is a fit to the world's data in the DIS region. The dotted line indicates the slope for the integral at $Q^2 = 0$ predicted using the GDH sum rule. The short-dashed line is the result from a calculation that takes into account the contribution from the nucleon resonances only (code AO) \cite{burkert}. The long-dashed line \cite{burkert2} is the AO result plus a term that depends smoothly on $Q^2$ and interpolates between the resonances and the GDH limit at $Q^2 = 0$. Fig.4 also shows the prediction from the model by Soffer and Teryaev \cite{soffer}(dot-dashed line). The solid triangle are the CLAS EG1 data alone while the open triangles include the estimated contribution to the integral from beyond our kinematics limits. The inner error bars are statistical and the outer error bars represent the systematic errors added in quadrature.

The first conclusion one can draw from Fig. 4 is that the integral over the measured region is in rather good agreement with the prediction of the AO parametrization for the resonance contribution only. Therefore, the non-resonant piece must contribute relatively little to the integral over the resonace region in the case of the deuteron (perhaps due to a partial cancellation between the asymmentry of the proton and the neutron).

In general, one can say that the integral over $g_1^d$ follows the expected trend, rising towards the DIS limit at the highest measured $Q^2$ while dropping rapidly below 0 towards the lowest $Q^2$ point. Clearly, neither the kinematic range reach nor the statistical precision of the present data set allows a definite statement about the validity of the GDH sum rule limit. However, new CLAS data with improved statistical precision and kinematic coverage will better constrain the theoretical predictions.

Results on the integral $\Gamma_1$ for the proton, neutron and the proton-neutron difference will be soon available.

\begin{figure}[htb]
\includegraphics[width=7.5cm, height=8.5cm]{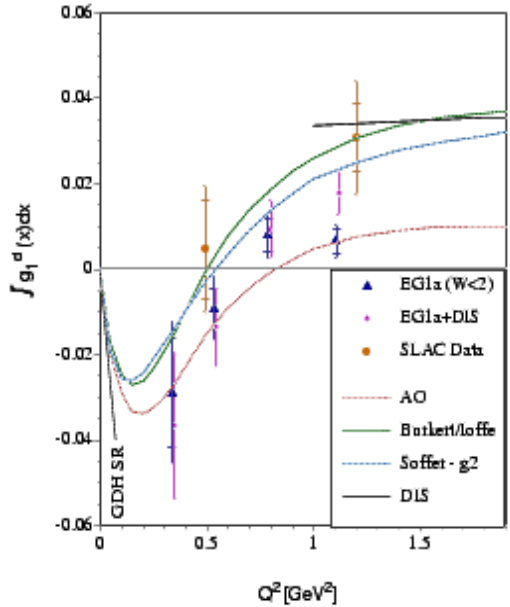}
\vspace{-1.8cm}
\vspace{.4cm}
\caption{The first moment of the spin structure function $g_1^d$ of the deuteron (per nucleon).}
\label{fig:fig4}
\end{figure}

\section{NUCLEON TOMOGRAPHY}
Much of our current knowledge of the nucleon comes from inclusive electron scattering experiments that measure elastic form factors and longitudinal parton densities. However, elastic scattering and deeply inelastic scattering give two orthogonal one-dimensional projections of the proton. The former measures the probability of finding a proton with a transverse size matching the resolution of the probe, while the latter probes the quark's longitudinal momentum distribution, but has no sensitivity to the transverse dimension. The information resulting from these two types of experiments is disconnected, and does not allow one to construct the image of a real 3-dimensional proton.

Semi-exclusive measurements, in which one hadron is observed in addition to the scattered electron, are needed to study the flavour structure of the nucleon and only fully exclusive processes in which all final products are reconstructed can unravel the complete internal dynamics.

The theoretical framework for the interpretation of these new class of experiments is given by the formalism of Generalized Parton Distributions (GPDs) \cite{muller,ji,rad} which give information on quark-quark correlations, transverse quark momentum distributions and contributions of correlated quark-antiquark pairs (mesons) to the nucleon wave function.

The basis for this approch are the ``handbag'' diagrams shown in Fig. 5. In fact, it has been shown that in the Bjorken scaling regime, the scattering amplitude for exclusive processes can be factorized into a hard-scattering part (exactly calculable in perturbative QCD) and a nucleon structure part, the lower blob of the diagrams, parametrized via GPDs. In addition to the dependence on the parton momentum fraction {\it x}, GPDs depend on two more parameters, the skewedness {\it $\xi$} (which in the Bjorken regime  is related to the momentum imbalance between the struck quark and the quark that is put back into the final state baryon) and the momentum transfer {\it t} to the baryonic system.

The complete extraction of the spin-dependent and spin-independent GPDs requires an extensive program involving measurements of a variety of channels and observables over a broad kinematic range.

\begin{figure}[h]
\begin{center}
\includegraphics[width=5.5cm, height=4.cm]{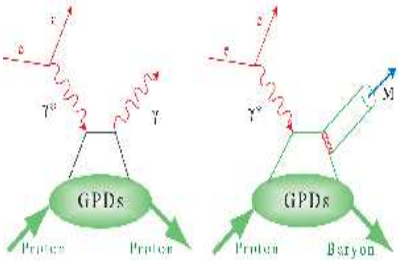}
\vspace{-0.5cm}
\caption{The ``handbag'' diagrams for GPDs.}
\label{fig:fig5}
\end{center}
\end{figure}
\subsection{Deeply Virtual Compton Scattering (DVCS)}
The simplest process that can be described in terms of GPDs is deeply virtual Compton scattering. In fact while exclusive meson production requires high energies and high photon virtualities to reach the Bjorken regime, DVCS may access the GPDs already at $Q^2$ as low as $1\ \rm{(GeV/c)^2}$ \cite{vander}.\\
One of the first experimental observation of DVCS has been obtained from the recent analysis of CLAS data with a 4.2 GeV polarized electron beam in a kinematical regime near $Q^2 = 1.5\ \rm{GeV^2} $ and $ x_b=0.22 $ \cite{step}.\\

\begin{figure}[h]
\vspace{4pt}
\begin{center}
\includegraphics[width=5.5cm, height=3.5cm]{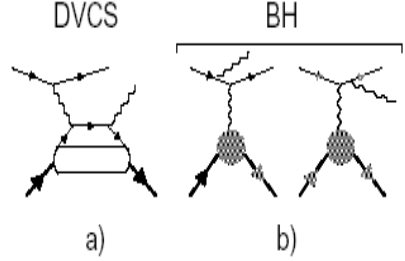}
\vspace{-1.5cm}
\vspace{.4cm}
\caption{Feynman diagrams for DVCS and Bethe-Heitler processes contributing to the amplitude of $ep\rightarrow e'p\gamma$ scattering.}
\label{fig:fig6}
\end{center}
\end{figure}

This experiment measures DVCS via the interference with the Bethe-Heitler process (BH) (Fig. 6b). For beam energies accessible at Jefferson Lab, the BH contribution in the cross section is several times larger than DVCS contribution, but this can be turned into an advantage by using a longitudinally polarized electron beam. In fact one can measure the helicity-dependent interference term that is proportional to the imaginary part of the DVCS amplitude. In this case the pure real BH contribution is subtracted out in the cross section difference.

For this analysis, electron and proton were both detected in the CLAS detector, the reaction $\vec{e}p \rightarrow epX$ was studied and the number of single photon final states was extracted by fitting the missing mass $(M_X^2)$ distributions. The beam spin asymmetry (BSA) was then calculated as:
\begin{equation}
BSA =\frac{1}{P_e}\frac{(N_\gamma^+ - N_\gamma^-)}{(N_\gamma^+ + N_\gamma^-)}
\end{equation}
where $P_e$ is the beam polarization and $N_\gamma^{+(-)}$ is the extracted number of  $\vec{e}p \rightarrow ep\gamma$ events at positive (negative) beam helicity. The resulting $\phi$ dependence is shown in Fig. 7.\\
A fit to the function
\begin{equation}
F(\phi) = A \sin(\phi) + B \sin(2\phi)
\end{equation}
yields $A = 0.217\pm0.031$ and $B = 0.027 \pm 0.022$
If the handbag diagram dominates, as expected in the Bjorken regime, B should vanish and only a contribution from the transverse photon should remain, as described by the parameter A \cite{die}.

The clear asymmetry, as expected from the interference of the DVCS and BH process, strongly supports expectations that DVCS will allow access to GPDs at relatively low energies and momentum transfers. Moreover the results agree in sign and are not far in magnitude from predictions based on available models of GPD parametrizations.

\begin{figure}[h]
\includegraphics[width=7.0cm, height=7.cm]{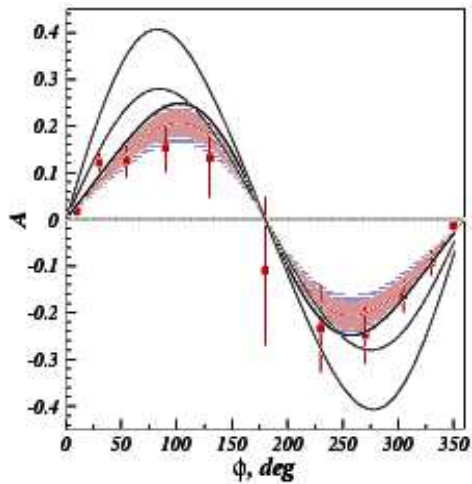}
\caption{$\phi$ dependence of the $\vec{e}p \rightarrow ep\gamma$ beam spin asymmetry at 4.25 GeV. The shaded region is the range of the fit function A($\phi$) defined by statistical and systematical uncertainties. The curves are model calculations according to \cite{beli}.}
\label{fig:fig7}
\end{figure}

\section{DYNAMICS OF THE STRONG INTERACTION}
One of the primary goal of nuclear physics is the study of the interplay between hadronic and partonic degrees of freedom, and of the effectiveness of traditional nucleon-nucleon theories or QCD inspired models in describing the data.
In this respect deuteron photodisintegration and photoproduction of light vector mesons at high momentum transfer {\it t} are two suitable reactions to study.
\subsection{Deuteron photodisintegration in the Quark-Hadron Picture}
Deuteron photodisintegration is well suited for studying 
nuclear reactions in the intermediate energy regime where neither the traditional meson exchange models nor pQCD describe the data well.

At high incident photon energy and intermediate angles, conventionally $90^\circ$, the $\gamma d \rightarrow pn$ differential cross section is well-described by the constituents counting rules (CCR) \cite{ccr}:
\[\frac{d\sigma}{dt} = \frac{1}{s^{n - 2}} f(\theta_{CM})\]
where $n$ is the minimum number of microscopic fields involved 
in the reaction, $s$ is the square of the total energy, and $\theta_{CM}$
is the proton scattering angle. 
In this case, $n = 13$, and then the CCR predicts
$d\sigma/dt \propto s^{-11}$.\\
Data on the deuteron photodisintegration differential cross section at large angles, $\theta_{CM} = 69^{\circ}$ and $89^{\circ}$, for $E_{\gamma} \geq 1$ and at $\theta_{CM} = 37^{\circ}$ and $53^{\circ}$ for 3 GeV and 4 GeV, respectively, follow the scaling prescription. In contradiction, polarization observables measured at $90^{\circ}$ and for photon energies up to 2 GeV are not consistently interpreted in a perturbative picture. Therefore, observed scaling in the cross section can not be taken as a stringent test that the perturbative regime as been reached. In fact, traditional models based on meson exchange current, are also able to reproduce this scaling law for $\theta^{CM}_p = 89^{\circ}$ \cite{AMEC}.  

In this context the use of non-perturbative QCD calculations to describe the deuteron photodisintegration process appears more suitable.

This is done by the so-called quark gluon string model (QGSM) \cite{gris}.
The reaction $\gamma d \rightarrow p n$ is described by the exchange 
of three valence quarks in the $t$-channel plus any number of gluons. This corresponds to the formation and break-up of a quark-gluon string in the intermediate state, leading to the factorization of the amplitudes. Such a string can also be identified with the nucleon Regge trajectory since the QGSM can be considered as a microscopic model for the Regge phenomenology, and can be used for the calculation of different quantities that have been considered before only at a phenomenological level \cite{kaida}.

Open questions include at what momentum transfer does one reach the perturbative regime, which is the most convenient description in the transition region, and where the conventional picture of the deuteron in terms of nucleons and mesons fail, and partons start to come into play.

The CLAS data on the complete angular distributions for the outgoing proton ($\theta_{LAB} = 10^{\circ}-140^{\circ}$) and for photon energies between 0.5 and 3.0 GeV contribute significantly to answering these questions.
Differential cross sections $d\sigma / d\Omega$ are reported in Fig.8 as a function of the proton angle in the CM frame, for fixed photon energy above 0.9 GeV and up to 2.45 GeV. The present preliminary results \cite{mirazita}, obtained from the analysis of about $30\%$ of the accumulated statistics (and with an additional cut $\theta_{CM} > 20^{\circ}$) show a persistent forward-backward asymmetry.
These data agree well with the previous available data from JLab \cite{tjnafexp} and SLAC \cite{slac2} in the region of overlap.

Also showed in Fig.8 are the QGSM calculations (solid line). 
The agreement with the data is very good and the angular dependance of the cross section at fixed photon energy is well reproduced included the observed forward-backward asymmetry. 

\begin{figure}[htb]
\includegraphics[width=8.5cm, height=11.cm]{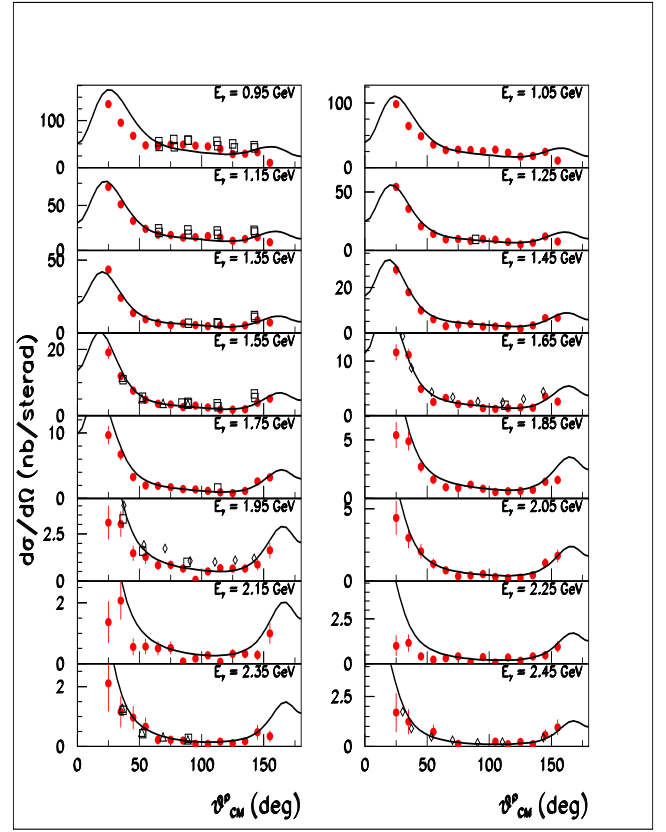}
\vspace{-1.cm}
\caption{Preliminary results of the $\gamma d \rightarrow pn$ differential cross section measured with CLAS (solid dots), compared with the published data from Jlab \cite{tjnafexp} (open triangles) and SLAC \cite{slac2} (open squares). The curve is the QGSM calculation \cite{gris}.}
\label{fig:fig8}
\end{figure}

\subsection{Photoproduction of the $\phi$, $\rho$ and $\omega$ mesons at large momentum transfer}
At low values of momentum transfer t, photoproduction of vector mesons occurs mainly through the photon coupling to intermediate vector meson states which diffractively scatter from the target. This corresponds to the Vector Meson Dominance Model (VDM), and the cross section depends only on the size of the meson and the target. At high t, hard processes are expected to take over and the production is thus more sensitive to quark and gluon exchange mechanisms.

CLAS experiments have measured the {\it t} dependence of the photoproduction of $\rho$, $\omega$ and $\phi$ on the proton up to values of {\it t} around 5 $\rm{GeV^2}$ where scarce data were available for $\rho$ and $\omega$ and no data existed for $\phi$ production for ${\it t} \geq 1\ \rm{GeV^2}$.

The results are shown in Fig. 9 \cite{anciant,bat1,bat2} where the $d\sigma / dt$ is plotted versus {\it t}. As expected, at small {\it t} the cross section is well described as a purely diffractive process in the framework of the traditional VDM, or in a modern way as the exchange of the pomeron trajectory in the {\it t} channel \cite{donnachie}. At larger {\it t}, the small impact parameter makes it possible for a quark in the vector meson and a quark in the proton to come close enough to exchange two gluons which do not have enough time to reinteract to form a pomeron. Large momentum transfer also select configurations in which the transverse distances between the two quarks in the vector meson and the three quarks in the proton are small. In that case, each gluon can couple to different quarks of the vector meson, as well as to different quarks of the proton. Because of the dominant {\it $s \bar{s}$} component of the $\phi$, and to the extent that the strangeness component of the nucleon is small, in $\phi$ photoproduction the exchange of quarks is strongly suppressed. This is clearly shown in Fig. 9 (upper panel) where above ${\it t} \approx 1.8\ \rm{GeV^2}$, the data rule out the diffractive Pomeron and the two-gluon realization alone (solid line) is able to reproduce the experimental data (except the last point at ${\it t} = 3.9\ \rm{GeV^2}$ where one approches the kinematical limit and {\it u} channel nucleon exchange may contribute \cite{laget}).

The two-gluon exchange mechanism alone, that fully describes the $\phi$ photoproduction data, instead, badly misses the cross section at large momentum transfer for the $\rho$ and $\omega$ photoproduction (Fig. 9 central and bottom panel). In this case, in the QCD inspired model of Ref.\cite{donnachie,laget,laget2}, a good agreement with the data is achieved when quark interchange processes are also included in the calculations. This hard-scattering mechanism is incorporated in an effective way by using the so-called saturated, {\it i.e.} non-linear, Regge trajectorie  \cite{serg}.
In conclusion, photoproduction cross section of the light vector mesons $\phi$, $\rho$, $\omega$, has a flat behaviour at large {\it t}. This feature is the evidence of the presence of hard processes well described in a QCD inspired model.

\begin{figure}[ht]
\includegraphics[width=6.7cm, height=5.6cm]{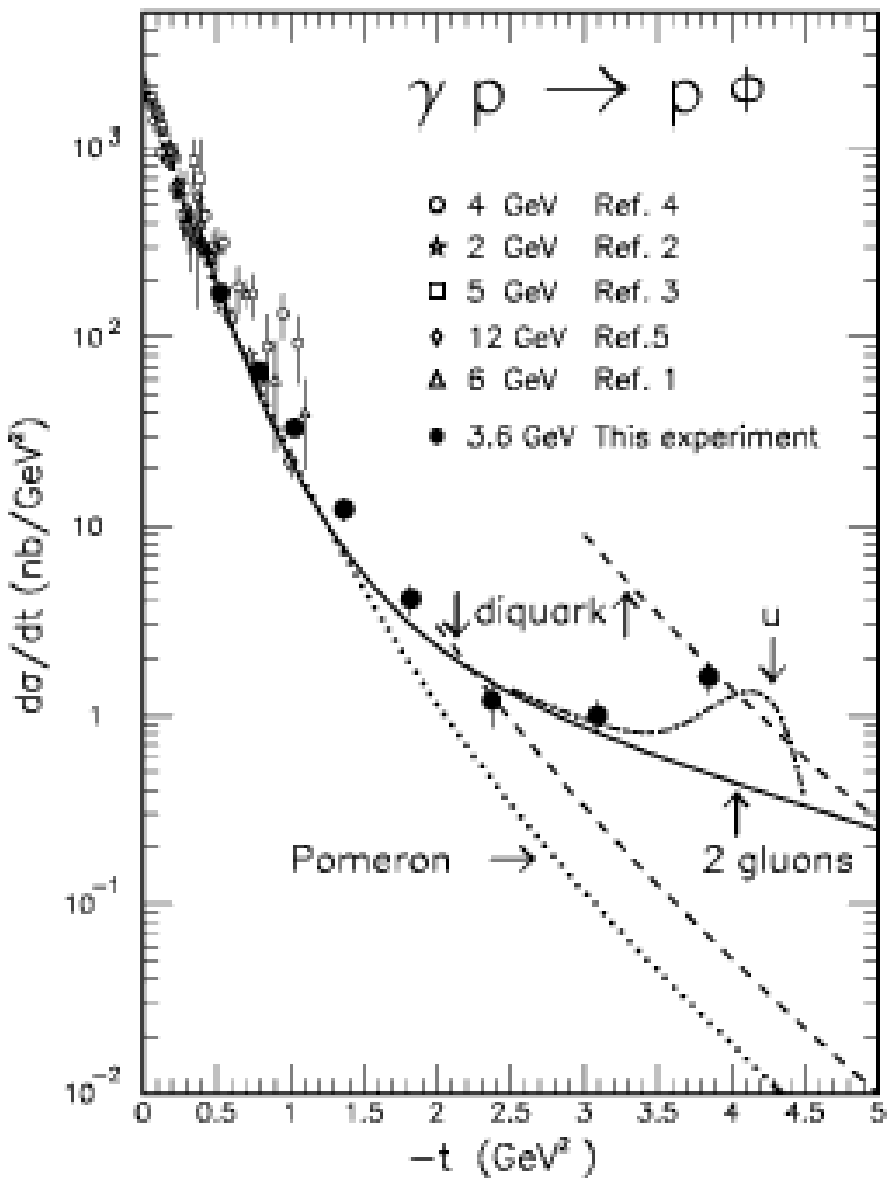}
\hspace*{0.20cm}
\includegraphics[width=6.22cm, height=5.6cm]{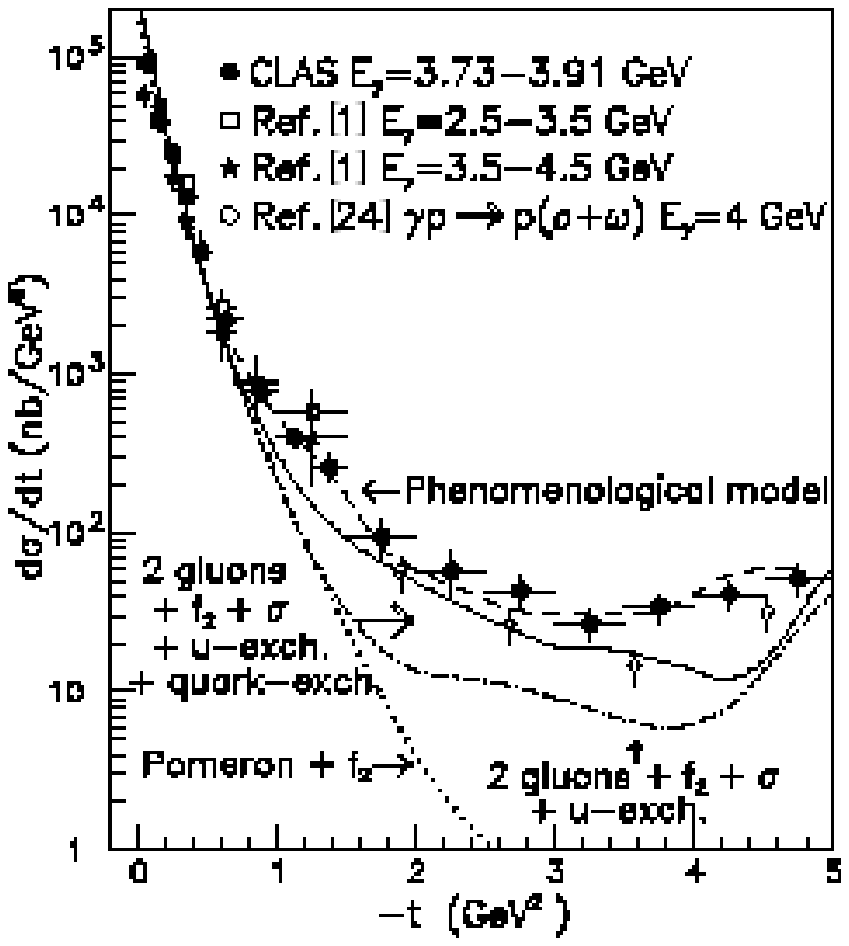}
\end{figure}

\begin{figure}[h!]
\vspace{-0.5cm}
\hspace*{0.1cm}
\includegraphics[width=7.2cm, height=5.5cm]{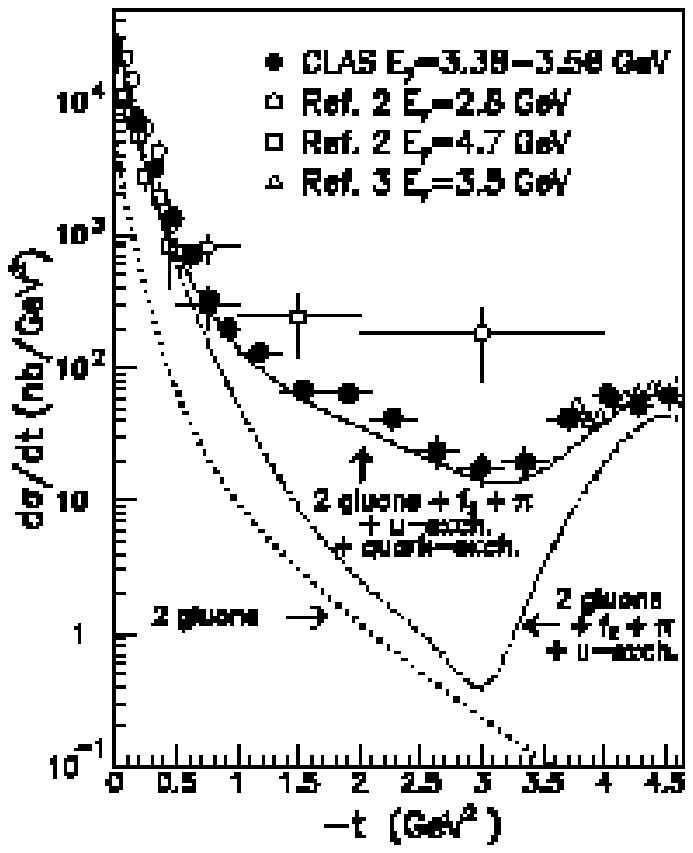}
\vspace{-0.9cm}
\caption{Differential cross sections for vector meson photoproduction at CLAS: $\phi$ (upper panel), $\rho$ (central panel) and $\omega$ (bottom panel). For the explanations of the curves see Ref. \cite{anciant,bat1,bat2}.}
\label{fig:fig9}
\end{figure}

\section{CONCLUSIONS}
The large and broad experimental effort of the CLAS collaboration at Jefferson Lab is providing a wealth of new data that will help in clarifying our understanding of nucleon structure and of nuclear dynamics in the intermediate energy region which is the domain of strong QCD. Only a few examples of these new results have been reported here, highlighting the main topics on which the experimental program is focused, {\it i.e.} electroproduction of nucleon resonances, spin structure functions and spin integrals $\Gamma_1$, new structure functions (GPDs) and analysis of the some ``hard'' processes ( deuteron photodisintegration and vector meson photoproduction). The new data put stronger constraints on QCD-inspired calculations and show a new avenue for the study of nucleon structure which is inaccessible in inclusive scattering experiments. Many more results are expected to come from the considerable amount of new data accumulated with CLAS.

\end{document}